# The Importance Analysis of Use Case Map with Markov Chains

Yaping Feng
School of Computer engineering, Kumoh National Institute of Technology, Korea
.

Lee-Sub Lee
School of Computer engineering, Kumoh National Institute of Technology, Korea
.

*Abstract*—UCMs (Use Case Maps) model describes functional requirements and high-level designs with causal paths superimposed on a structure of components. It could provide useful resources for software acceptance testing. However until now statistical testing technologies for large scale software is not considered yet in UCMs model. Thus if one applies UCMs model to a large scale software using traditional coverage-based exhaustive tasting, then it requires too much costs for the quality assurance. Therefore this paper proposes an importance analysis of UCMs model with Markov chains. With this approach not only highly frequently used usage scenarios but also important objects such as components, responsibilities, stubs and plug-ins can also be identified from UCMs specifications. Therefore careful analysis, design, implementation and efficient testing could be possible with the importance of scenarios and objects during the full software life cycle. Consequently product reliability can be obtained with low costs. This paper includes an importance analysis method that identifies important scenarios and objects and a case study to illustrate the applicability of the proposed approach.

*Keywords- Use Case Maps; Markov chain; Usability Testing.*

I. INTRODUCTION

UCMs (Use Case Maps) [1, 2, 3] is a set of semi-formal notations for describing scenarios of a system. This notation is being standardized as a part of the URN (User Requirement Notation) that is the most recent addition to ITU-T's (International Telecommunication Union-Telecommunication) family of languages. UCMs model provides a scenario-based model for a system. And it has been successfully applied to wide range of systems, including telecommunication systems [4], distributed systems [1] and etc. Furthermore, Daniel Amyot applied UCMs model for customer-oriented acceptance tests for Web applications [5]. But existing research works have not considered statistical testing method yet. Traditional coverage-based exhaustive testing might be impractical for the large scale software products because of the size as well as the generally uneven distribution of problems and usage frequencies in different areas and product components.

The general solution of testing for large software products uses product reliability goals as a stopping criterion. The use of the criterion requires testing to be performed under an environment that resembles actual usage by target customers so that realistic reliability assessment can be obtained, resulting in the so-called SUT (Statistical Usage Testing) [3].

A prerequisite to such statistical testing and reliability analysis strategies is collection of usage information and construction of corresponding usage models. UCMs describe functional requirements and high-level designs with causal scenarios superimposed on a structure of components. Furthermore individual scenarios from UCMs can be generated automatically from related tools [6]. Hence, it is straightforward to collect usage information from UCMs to construct usage model.

This paper proposes an importance analysis of UCMs model with Markov chains which applies statistical technique to UCMs model. If one simply applies traditional statistical techniques to usecase model, only highly frequently used usage scenarios can be identified and tested more thoroughly than less frequently used ones. But the approach of this paper is based on UCMs model which includes more objects like components, responsibilities, stubs and plug-ins. So this paper has proposed mechanisms and object model of UCMs which can calculate the importance of these objects. Thus efficient testing can be given to both highly frequently used usage scenarios and important objects to assure and maximize the product reliability from a customer's perspective with low cost. Since the combination of UCMs model and statistical technique can identify much more important information, it will be very beneficial for quality assurance engineering.

This paper is organized as follows. Related works are discussed in section 2. Section 3 describes the background concepts of UCM scenario models and Markov chain usage models. An importance analysis of UCM with Markov chains is presented in section 4 with a case study. Conclusions and future works are discussed in section 5.

II. RELATED WORKS

Markov chain usage model is a probabilistic model based on FSM (Finite State Machine) model. It can be generated from textual use cases [7, 8] and then the most likely usage scenarios are identified as test cases. Since these works are based on the normal use cases, with these traditional approaches important components and events can not be identified.

Kallepalli and Tian developed UMMs (Unified Markov Models) to support statistical testing, performance evaluation,

This paper was supported by Research Fund, Kumoh National Institute of Technology

and reliability analysis [9]. They applied the model to web application testing and proposed an automated way of constructing UMMs model from web server logs [9, 10]. With this approach, one web page or one information file serves as a state, so components such as java components and events cannot be included in the scenarios. Due to the complexity of the model level in the web applications, identifying important components and events did not be covered.

Daniel Amyot applied UCM model to customer-oriented acceptance tests for Web applications [5]. But complete coverage of acceptance tests in this approach is impractical for the large projects. It is necessary to identify highly frequently used usage scenarios and components in order to achieve efficient tests to assure and maximize the product reliability.

III. BACKGROUND

*A. Use Case Map*

UCM is part of user requirements notation standards proposed to the ITU-T for describing functional requirements as causal scenarios. UCM allows developers to model dynamic behavior of systems where scenarios and structures may change at run-time. For these reasons, UCM has been widely used in a range of systems [3].

UCM notation consists of three formal elements that form the basis for all maps: paths represent scenarios; components including system and non-system entities perform responsibilities; and responsibilities such as actions, events and so on are linked to paths and may be contained within components to indicate that the component executes that event. UCM scenarios begin with start points which represent triggering events and/or pre-conditions which are required for the commencement of the scenario. Scenario paths are followed to endpoints that represent terminating events or post conditions of the scenario execution. The UCM notation also provides a hierarchical abstraction mechanism in the form of stubs (diamonds) and plug-ins (sub-maps). Each hierarchy of maps has a root map that contains stubs where lower-level maps can be plugged in.

UCM are primarily visual, but a formal textual representation also exists. Based on the XML (eXtended Markup Language) 1.0 standard, this representation allows for tools to generate UCMs or use them for further processing and analysis [11].

We mentioned Figure 1 to demonstrate the fundamental understanding of UCMS and it will be used as an example in the rest of the paper. The figure shows a simple UCMs model where UserO attempts to establish a telephone call with another UserT through some network of agents. Each user has an agent responsible for managing subscribed telephony features. UserO first sends a connection request (req) to the network through his agent. UserO's agent has an originating dynamic stub SO which has two plug-ins, default originating and OCS. UserO can subscribe to one of services. The OCS plug-in shows an object OCSlist that represents a list of screened numbers that the originating user (UserO) is forbidden to contact. The called number is checked against the list (chk). If the call is denied, a relevant message is prepared for the originating party (md). This successful request causes the called agent to verify (vrfy) whether the called party is idle or busy (terminating plug-in in stub ST). If he is idle, then there will be some status update (upd) and a ring signal will be activated on UserT's side (ring), concurrently, preparation of a ring-back targeted to the originating party (mrb). Otherwise, a message stating that UserT is not available will be prepared (mb) and sent back to UserO (msg). For a detailed description of UCMs notation, readers should refer to [2] and [12].

Compared to the traditional use case diagrams, UCMs contain more object information such as components, responsibilities, stubs and plug-ins. Thus this paper applies statistical technique to UCMs model which can analyze not only scenarios but also these additional objects to assure and maximize the product reliability from a customer's perspective.

*B. Markov chain usage models*

Markov chains is the simplification based on the so called memoryless property, or the Markovian property, which states that the state transitions from a given state depend only on the current state, but not the history or how we reached that particular state [13].

As shown in Figure 2, a usage chain consists of states, The Markov chain is completely defined when transition probabilities are established that represent the best estimate of real usage. For each state the summation of output arcs or probabilities should be 1. Transition probabilities can be obtained from various sources containing information about the actual usage counts and relative frequencies. Several methods can be employed to extract this information, including subjective evaluation based on expert opinions, survey of target customers, and measurement of actual usage logs.

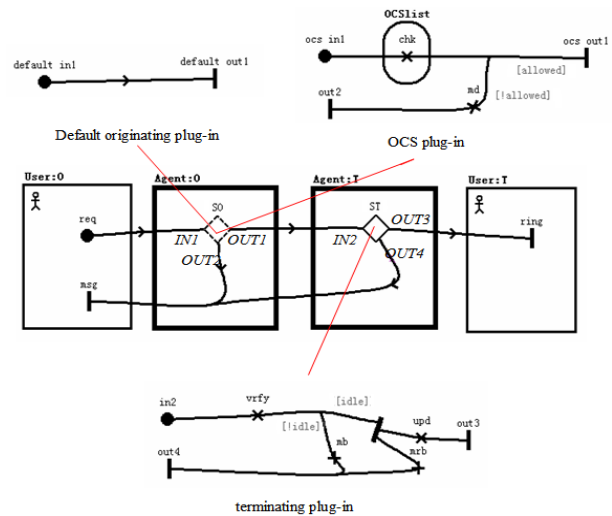

Figure 1. A Simple telephone system UCMs model

An operational sequence consisting of visits to multiple states can be constructed by following the state transitions. The likelihood for a particular sequence to happen can also be easily calculated by the product of its individual state transition probabilities. Therefore, Markov chains can be used to ensure performance and reliability based on usage scenarios and frequencies by target customers. Furthermore, Kallepalli and Tian developed UMMs to support statistical testing, performance evaluation, and reliability analysis. UMMs possess a hierarchical structure formed by a collection of Markov chains. It captures information about execution flow (control flow), transaction processing (workload creation, handling, and termination), and associated probabilistic usage information.

As introduced in section 3.1, UCMs model also has a hierarchical structure, a root map that contains stubs where lower-level maps can be plugged in. Thus it is straightforward to convert UCMs model to UMMs.

For example as shown in Figure 3, the top level Markov chain, depicted by the upper part of the diagram, represents high-level operational units (states), associated connections (transitions), and usage probabilities. Various sub operations may be associated with an individual state and can be modeled by more detailed models or sub models, as shown in the down Markov chains.

IV. AN IMPORTANCE ANALYSIS OF UCM WITH MARKOV CHAINS

This paper proposes an importance analysis of UCMs model with Markov chains which can identify highly frequently used usage scenarios and generate lists of important objects such as components, responsibilities, stubs and plug-ins in UCMs specifications. This section will introduce overall phases and detail procedure with an example.

The main procedure of this approach is illustrated in Figure 4.

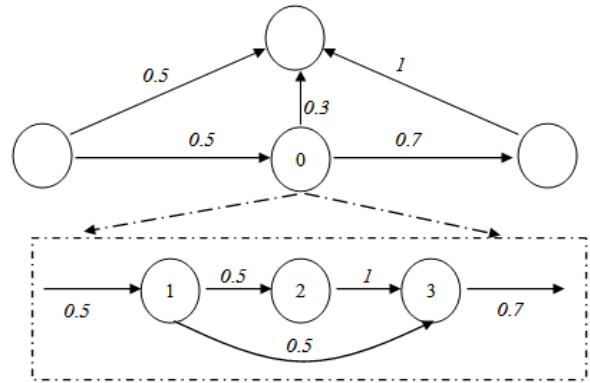

Figure 3. UMMs example

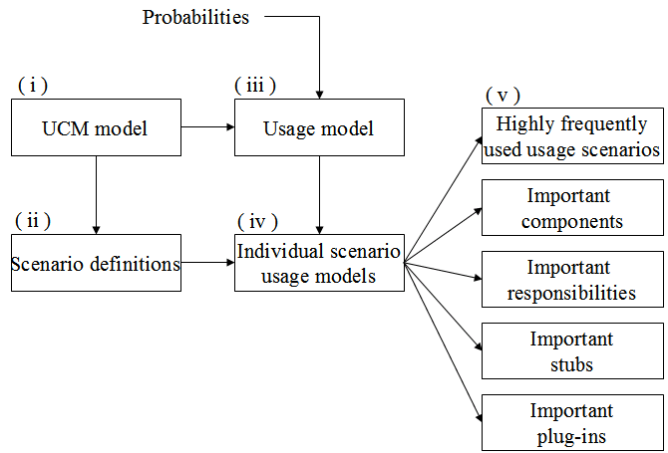

Figure 4. Overall procedure of the proposed approach

This approach includes five steps as follows:

*1)* Construct a UCMs model during requirement phase: A UCMs model should be constructed during requirement phase (i); UCMNav or jUCMNav tool can be used to draw UCM notations [3].

*2)* Define all the possible scenarios from the UCMs model: Since it is easier to identify transition probability on scenario basis; the next step is generating all of the possible scenarios from the UCMs model. The generation of scenarios is for Transition probabilities should be prepared for each scenarios path

*3)* Convert the UCMs model to an usage model with probabilities: After assigning probabilities, the UCMs model is converted to a usage model based on some rules that will be discussed later

*4)* Generate each individual scenario usage model.

*5)* An importance analysis of the objects of each object types of the UCMs: With the individual scenario usage models, an importance analysis should be followed to identify highly frequently used usage scenarios and generate lists of objects (components, responsibilities, stubs and plug-ins) ranked by their relative importance factor in the early phases of software life cycle.

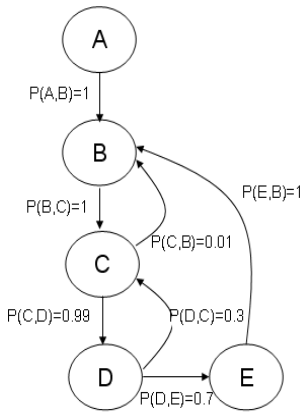

Figure 2. A Markov usage chain example

TABLE I. SCENARIO DEFINITIONS FOR SIMPLE TELEPHONE SYSTEM

| Scenario name | Pre-condition | Post-condition |
|---|---|---|
| NormalIdleCall | Start point = req<br>SO = default in1<br>ST = in2<br>[idle] = true | null |
| NormalBusyCall | Start point = req<br>SO = default in1<br>ST = in2<br>[idle] = false | null |
| OCSDeniedCall | Start point = req<br>SO = ocs in1<br>[idle] = false | null |
| OCSAllowedIdleCall | Start point = req<br>SO = ocs in1<br>ST = in2<br>[allowed] = true<br>[idle] = true | null |
| OCSAllowedBusyCall | Start point = req<br>SO = ocs in1<br>ST = in2<br>[allowed] = true<br>[idle] = false | null |

TABLE II. THE CORRESPONDENCES BETWEEN UCMs MODEL AND USAGE MODEL

| UCMs model | Usage Model |
|---|---|
| Start/end points | States |
| Responsibilities | States |
| Scenario path | Transition path |
| AND/OR forks and joins | AND/OR forks and joins |

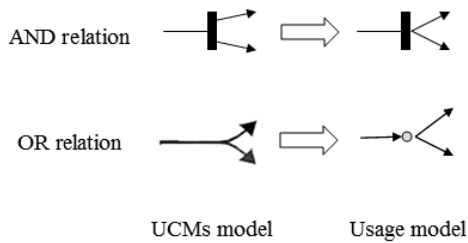

Figure 5. "AND/OR" relation translation

### B. An UCMs model construction

UCMs graphical models describe functional requirements and high-level designs with causally linked responsibilities, superimposed on structures of components. With the help of editor UCMNav or jUCMNav, it is easy to construct UCMs model. A detailed introduction of how to use these editors can be found in [3].

For an example, a UCMs model of a simple telephone system is illustrated in Figure 1. It is designed with jUCMNav editor.

### C. Identifying Scenario definitions for UCMs model

For extracting individual scenarios, all possible scenarios should be defined. For each scenario, one needs to provide a scenario name, pre-condition including the list of start points to be triggered and initial values of global Boolean variables, and (optionally) a post-condition used to assert the validity of a scenario once the traversal has completed.

As shown in Figure 1, a simple telephone system is modeled with jUCMNav editor. The following five scenarios are identified from the UCMs model as shown in Table 1:

*1) NormalIdleCall:* UserO subscribes default originating plug-in and gives a call to UserT who is idle.

*2) NormalBusyCall:* UserO subscribes default originating plug-in and gives a call to UserT who is busy.

*3) OCSDeniedCall:* UserO subscribes originating call screening plug-in and gives a call to UserT, but this number is in the list of screened numbers that the originating UserO is forbidden to contact.

*4) OCSAllowedIdleCall:* UserO subscribes originating call screening plug-in and gives a call to UserT whose number is allowed and UserT is idle.

*5) OCSAllowedBusyCall:* UserO subscribes originating call screening plug-in and gives a call to UserT whose number is allowed and UserT is busy.

### D. Converting UCMs model to usage model with probabilities

The general approach of usage model construction includes two steps. First, construct a basic model with basic states (nodes) and state transitions (links) identified from product specification. Second, complete the usage model by assigning transition probabilities.

There are some quite close correspondences between some of the scenario entities in UCMs model and usage model elements. Table 2 illustrates these correspondences. In the Table 2, UCM start/end points or responsibilities can represent states in usage model. Causal relationships path between responsibilities in UCMs model can serve as transition path between states in usage model. AND/OR forks and joins relationships serve as AND/OR forks and joins in usage model. The AND/OR relation is shown in Figure 5.

"Stub" elements in UCM root map can be converted to states in top-level models of UMMs. "Plug-in" elements in UCM sub-map can be converted to sub-level models of UMMs as shown in Figure 6. The big round with "S" means a state. The "P1" and "P2" identify the path1 and path2. If the "Stub" is a dynamic stub, an additional "or" relation by a small round notation is added.

The second step of the construction of a usage model is assigning transition probabilities. Several methods such as expert opinions, survey of target customers, and measurement of actual usage logs can be employed to extract these probabilities [10]. Basic UCMs model is augmented with

(IJCSIS) International Journal of Computer Science and Information Security,
Vol. 7, No. 1, 2010

assigning probabilities to start points, including plug-ins' start points, and branches on OR-forks.

As shown in Figure 1, a simple telephone system has been modeled by UCM. This UCMs model should be converted to usage model that is illustrated in Figure 7. This usage model is an UMM, the top-level UMM represents the root map in the UCMs model, and there are 2 sub-level UMMs which represent plug-ins of two stubs in top-level model. The numbers assigned to transitions represent transition probabilities.

### E. Individual scenario usage model generation

As shown in Table 1, five scenarios are defined for the example simple telephone system. Thus based on the whole usage model, individual scenario usage models can be generated. Figure 8 shows NormalIdleCall scenario usage model. Other four scenarios usage models can also be generated with the same way.

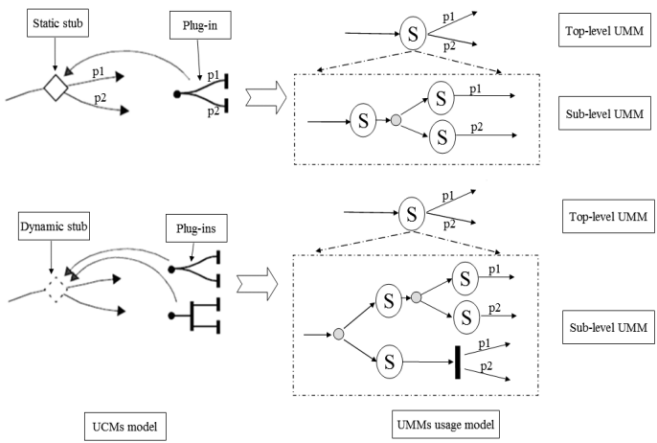

Figure 6.  "stub" and "plug-in" elements of UCMs in UMMs

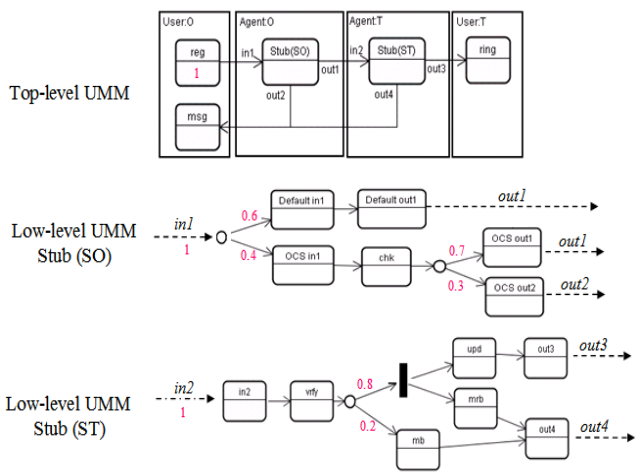

Figure 7.  Usage model for the simple telephone system

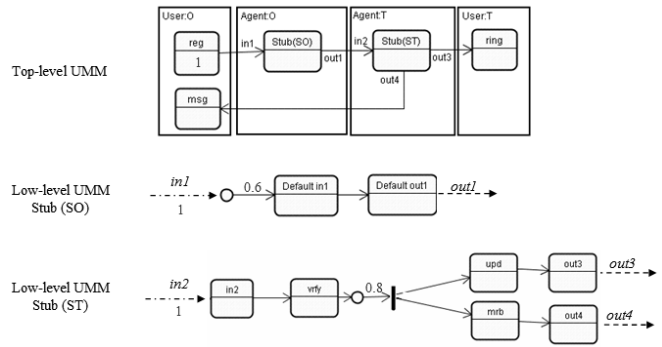

Figure 8.  NormalIdleCall scenario usage model

### F. An Importance Analysis of UCMs

In this section, an importance analysis methodology of UCMs model will be introduced. Firstly we will provide several definition of the model. Secondly with an example the insight of the equations for importance will be illustrated. Finally the equations are presented.

The definition of importance of an object is given as Definition 1. According to the definition, if an object is used more frequently by users, the importance of the object is higher.

**Definition 1:** *Importance of an object is usage probability of the object from a customer's perspective.*

As shown in the background of UCMs, there are several kinds of objects existed in UCMs model such as scenarios, components, responsibilities (start/end points), stubs and plug-ins. According to the description of UCMs, a component can contain responsibilities, stubs, plug-ins, and other components. And also a plug-in can contain responsibilities, stubs and components. We can easily notice that is has a composite pattern. Therefore all the objects in the UCMs model are classified as three types: scenarios, primitive objects and containers. The definitions of these objects are given as follows:

**Definition 2:** *A scenario is a path which consists of transitions from start point to end point across the related responsibilities which may bind to the components*.

**Definition 3:** *A primitive object is an object which is not includes any other objects*.

**Definition 4:** *A container is an object that consists of other objects such as primitive objects* or containers.

The importance of a scenario can be calculated as product of related individual transition importance or probability. For a primitive object, the importance is summation of product of each related scenario's importance and the number of appearances of primitive object in this scenario. The importance of a container can be calculated as summation of the importance of its contained objects.

Figure 9 shows an example of an object model to illustrate the insight of the calculation of the importance of each type of the object. This object model is actually a lattice model. In this example, four scenarios ($S_1$, $S_2$, $S_3$ and $S_4$) are defined, and

*(IJCSIS) International Journal of Computer Science and Information Security,*
*Vol. 7, No. 1, 2010*there are six primitive objects and four containers. Scenario $S_1$ has three transitions $T_1$, $T_3$ and $T_5$. Because the transition probabilities should be prepared initially, the importance of scenarios S1 can be calculated as $I(S_1) = I(T_1) * I(T_3) * I(T_5)$. The primitive object R01 appears in scenario $S_1$ once and in scenario $S_2$ also once. So the importance of primitive object R01 can be calculated as $I(R01) = I(S_1) * 1 + I(S_2) * 1$. For the container C03, it contains two child objects R01 and R03. Thus the importance of the container C03 is calculated as $I(C03) = I(R01) + I(R03)$. By this way, the importance of all the objects can be calculated.

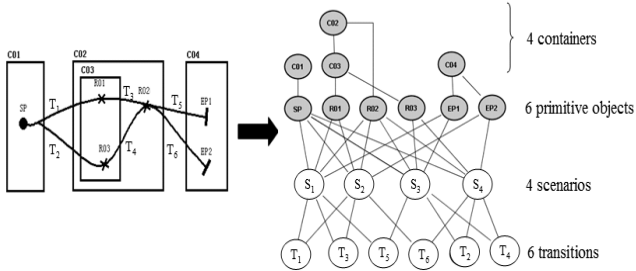

Figure 9. An example of object model for importance analysis of UCM

After generalizing the above example we can define formal equations of calculating the importance of all the objects in UCMs model as follows. The Eq.1 is used for the importance of a scenario. The Eq.2 is for the importance of a primitive object and the Eq.3 is used for a container.

$$I(S) = \prod_{T_i \in S} I(T_i) \qquad (1)$$

Where:

S represents a scenario

I(S) is importance of a scenario S

$T_i$ is a transition included in scenario *S*

$I(T_i)$ is importance of a transition $T_i$ in this scenario which is initially defined

$$I(PO) = \sum_{S_i \in S(C)} \big(I(S_i) \times N(PO, S_i)\big) \qquad (2)$$

Where:

PO represents a primitive object

I(PO) is importance of a primitive object PO

$S_i$ represents a scenario

$I(S_i)$ is importance of scenario $S_i$

S(C) is a set of scenarios which includes the primitive object PO

N(PO, $S_i$) is the number of the primitive object PO appeared in scenario $S_i$

$$I(C) = \sum_{O_i \in O} I(O_i) \qquad (3)$$

TABLE III. SCENARIOS IMPORTANCE FOR THE SIMPLE TELEPHONE SYSTEM

| Scenarios | Importance |
|---|---|
| NormalIdleCall | 0.48 |
| NormalBusyCall | 0.12 |
| OCSDeniedCall | 0.12 |
| OCSAllowedIdleCall | 0.224 |
| OCSAllowedBusyCall | 0.056 |

TABLE IV. IMPORTANT SCENARIOS WITH OVERALL IMPORTANCE THRESHOLD 0.2

| Scenarios | Importance |
|---|---|
| NormalIdleCall | 0.48 |
| OCSAllowedIdleCall | 0.224 |

TABLE V. TABLE 5. IMPORTANT SCENARIOS WITH ALTERNATIVE IMPORTANCE THRESHOLD 0.3

| Scenarios | Importance |
|---|---|
| NormalIdleCall | 0.6*0.8=0.48 |
| OCSDeniedCall | 0.4*0.3=0.12 |
| OCSAllowedIdleCall | 0.4*0.7*0.8=0.224 |

Where:

C represents a container

I(C) is importance of a container C

O(C) is a set of child objects of container C.

$O_i$ represents a contained object in the container C. The contained objects may be child primitive objects or containers

$I(O_i)$ is importance of object $O_i$

### G. Case study of the importance analysis of UCM: a simple telephone system

In this sub-section we will show an example according to the importance analysis method explained in 4.5. For the simple telephone system, five scenarios are identified from the UCMs model (Table 1). Thus by Eq. (1), the importance of these five scenarios is calculated and the result is shown in Table 3.

After calculating importance of scenarios, to identify the highly frequently used usage scenarios, some thresholds are used. In practical applications, thresholds can be adjusted to control the numbers of test cases to be generated and executed. Several thresholds have been initially proposed in [14] and used in developing UMMs. Two kinds of thresholds are used in this paper:

*1) Overall importance threshold for complete end-to-end operations to ensure that commonly used complete operation sequences by target customers are covered and adequately tested.*

*2) Alternative importance threshold to ensure commonly used operation pairs, their interconnections and interfaces are covered and adequately tested.*

*(IJCSIS) International Journal of Computer Science and Information Security,
Vol. 7, No. 1, 2010*TABLE VI. IMPORTANCE OF ALL RESPONSIBILITIES IN SIMPLE TELEPHONE SYSTEM

| Responsibilities | Importance |
|---|---|
| req | 1 |
| msg | 1 |
| vrfy | 0.88 |
| out4 | 0.88 |
| in2 | 0.88 |
| upd | 0.704 |
| ring | 0.704 |
| out3 | 0.704 |
| mrb | 0.704 |
| default out1 | 0.6 |
| default in1 | 0.6 |
| ocs in1 | 0.4 |
| chk | 0.4 |
| ocs out1 | 0.28 |
| mb | 0.176 |
| out2 | 0.12 |
| md | 0.12 |

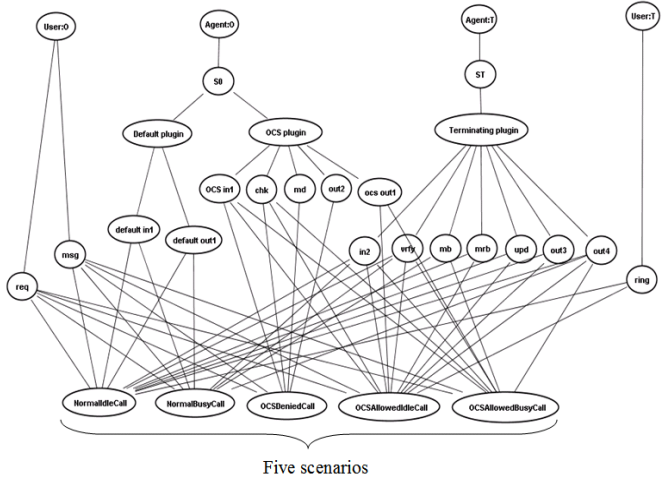

Figure 10. Composition Object model of UCMs model for simple telephone system

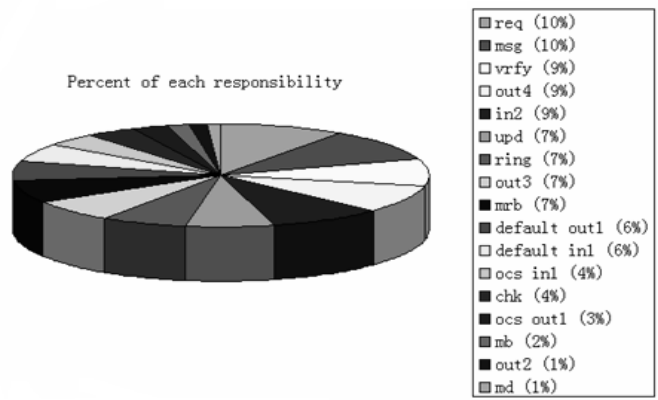

Figure 11. Percent of each responsibility importance

So if overall importance threshold is given as 0.2, then the highly frequently used usage scenarios can be identified as shown in Table 4. If alternative importance threshold is given as 0.3, Table 5 shows the highly frequently used usage scenarios.

For calculating the importance of components, responsibilities, stubs and plug-ins in UCMs model of simple telephone system, firstly a composition object model as shown in Figure 10 is given. This object model clearly describes the architecture of relations among different objects in UCMs model. Based on this object model, importance of each object can be easily calculated.

Responsibilities in this object model are all primitive objects, so with Eq. (2), the importance of all the responsibilities can be calculated as shown in Table 6.

Plug-ins in this object model are all containers, importance of each plug-in can be calculated as summation of its children's importance by Eq. (3). The result is shown as follows:

- Importance of plug-in Default ORIGINATING = 1.2
- Importance of plug-in TERMINATING = 4.928
- Importance of plug-in OCS = 1.32

Importance of each stub and component also can be calculated with Eq. (3). The result is shown as follows:

- Importance of stub SO = 2.52
- Importance of stub ST = 4.928
- Importance of component AgentT = 4.928
- Importance of component AgentO = 4.928
- Importance of component UserO = 2.52
- Importance of component UserT = 0.704

After getting all objects importance, a percent of importance for each object type can be calculated to show which objects are more important clearly. Percent of responsibility importance is shown in Figure 11.

The calculation of the importance is very simple and straightforward. As presented above the importance of each scenario can be easily calculated by Eq. (1) using transition probabilities, and after getting the importance of scenarios, the importance of primitive objects can be calculated with Eq. (2) based on the object model. For containers, the importance can be calculated as simple summation of its children's importance by Eq. (3).

In the early phases of the software life cycle (requirement analysis phase) the importance of all scenarios, components, responsibilities, stubs and plug-ins, and the highly frequently used usage scenarios and important objects can be identified from a customer's perspective. Since the importance of each object type can be identified at the early phase where UCM is defined, the quality engineering efforts can be applied from first phase to the last phase. That means this approach can support in the full software life cycle for quality assurance activities. For instance as shown in Figure 12, since the importance of each objects is identified at the Requirement Analysis phase, the important scenarios can be heavily tested on the Acceptance test, the important containers are can be took cared at the system testing and integration testing and the important primitive object can be focused at unit testing.

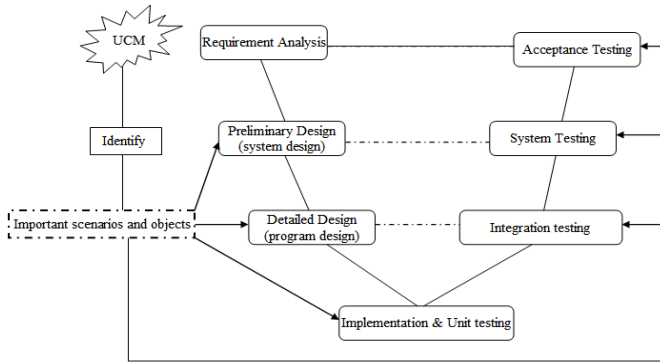

Figure 12. UCM in V model

## V. CONCLUSION

This paper proposes an importance analysis of UCM with Markov chains which applies statistical technique to UCMs model. Like other traditional statistical techniques, highly frequently used usage scenarios can be identified.

Furthermore, since the approach of this paper is based on UCMs model, it can identify more important object types such as components, responsibilities, stubs and plug-ins. Therefore this paper also proposes an object model of UCM which can calculate the importance of these objects. Thus the efficient testing can be possible for both highly frequently used usage scenarios and important object types to assure and maximize the product reliability from a customer's perspective. Since the combination of UCMs model and statistical technique can identify much more important information, it is very beneficial for quality assurance engineering.

The importance of each object is identified at the Requirement Analysis phase, which means the approach make possible the quality driven development during the full software life cycle. The method of the importance calculation is very clear, easy and straightforward that it is very practical to the real environment. To show these characters we include a simple telephone example with importance calculations.

In this approach, it requires manual operation to generate usage model and object models. Future work will extend the UCM Navigator (UCMNav or jUCMNav) editor to support automatic generation. Future work will also include the research of an algorithm to generate all possible scenarios automatically.

REFERENCES

[1] R.J.A. Buhr, "Use Case Maps as Architectural Entities for Complex Systems", IEEE Transactions on Software Engineering, Special Issue on Scenario Management. December 1998, Vol. 24, No. 12, pp. 1131-1155.

[2] "Use Case Maps Web Page and UCM User Group", http://www.UseCaseMaps.org, 1999.

[3] J.A. Whittaker, and M.G. Thomason, "A Markov chain model for statistical software testing", IEEE Trans on Software Engineering, Oct. 1994,Vol. 20, No. 10, pp. 812–824.

[4] Amyot, D., Specification and Validation of Telecommunications System wtih Use Case Maps and *Lotos*, in *School of Technology and Engineering*. 2001, University of Ottawa: Ottawa.d

[5] D.Amyot, J.F. Roy, and M.Weiss, "UCM-Driven Testing of Web Applications", 12th SDL Forum, June 2005, pp. 247-264.

[6] D. Amyot, X.Y. He, Y. He and D.Y. Cho, "Generating scenarios from use case map specifications", Third International Conference On Quality Software (QSIC), Dallas, USA, Nov. 2003, pp. 108–115.

[7] M. HÜBNER, I. PHILIPPOW, and M. RIEBISCH, "Statistical Usage Testing Based on UML", Proceedings of the 13th Annual IEEE International Symposium and Workshop on Engineering of Computer Based Systems, 2006.

[8] B. Regnell, P. Runeson, and C. Wohlin, "Towards Integration of Use Case Modelling and Usage-Based Testing", Journal of Systems and Software, 2000, pp. 50:117-130.

[9] C. Kallepalli, and J. Tian, "Measuring and Modeling Usage and Reliability for Statistical Web Testing", IEEE Transactions on Software Engineering, Nov. 2001, vol. 27, no. 11, pp. 1023-1036.

[10] C. Kallepalli, and J. Tian, "Usage Measurement for Statistical Web Testing and Reliability Analysis", Proceeding of Seventh International Software Metrics Symposium, London, IEEE Computer Society Press, 2001, pp. 148-158.

[11] D. Amyot, and A. Miga, Use Case Maps Linear Form in XML, version 0.13, May 1999.

[12] R.J.A. Buhr, and R.S. Casselman, "Use Case Maps for Object-Oriented Systems", Prentice-Hall, USA, 1995.

[13] S. Karlin, and H. M. Taylor, A First Course in Stochastic Processes, 2nd Ed. Academic Press, New York, 1975.

[14] J. D. Musa. Software Reliability Engineering. McGraw-Hill, New York, 1998.

AUTHORS PROFILE

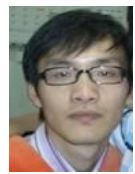

Yaping Feng received his B.S. degree in softtware engineering from East China Normal University, China in 2005 and received his M.S degree in computer engineering from Kumoh National Institute of Technology, Korea in 2007. He has worked as a developer in POSData China from 2008. His current research interests include Software Engineering and Mobile Computing.

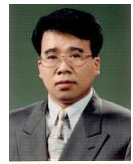

Lee-Sub Lee received his B.S., M.S. degrees in mathematics and computer science from Sogang University, Seoul, Korea, in 1988 and 1990, respectively. He ha a Ph.D degree in computer science & engineering from Korea University, Seoul, Korea in 2004. He is assistant professor of Kumoh National Institute of Technology since 2004. He had worked as a senior manager of the IT R&D Center, Samsung SDS, Ltd from 1990 in SungNam, Korea. His current research interests include Software Engineering and Mobile Computing.